\documentclass{elsart}
\usepackage{amsmath}
\usepackage{txfonts}
\usepackage{amsfonts}
\usepackage{amssymb}

\begin{document}
\begin{frontmatter}

\title{Black hole hair in higher dimensions}

\author{Chao Cao},
\ead{ccldyq@gmail.com}
\author{Yi-Xin Chen\corauthref{cor}},
\ead{yxchen@zimp.zju.edu.cn} \corauth[cor]{Corresponding author.}
\author{Jian-Long Li}
\ead{marryrene@gmail.com}

\address{Zhejiang Institute of Modern Physics}
\address{Zhejiang University, Hangzhou 310027, China}

\begin{abstract}
We study the property of  matter in equilibrium with a static,
spherically symmetric black hole in D-dimensional spacetime. It
requires this kind of matter has an equation of state \(\omega\equiv
p_r/\rho=-1/(1+2kn), k,n\in \mathbb{N}\), which seems to be
independent of D. However, when we associate this with specific
models, some interesting limits on space could be found:
(i)\(D=2+2kn\) while the black hole is surrounded by cosmic strings;
 (ii)the black hole can be surrounded by linear dilaton field only in
4-dimensional spacetime. In both cases, D=4 is special.

\end{abstract}


\end{frontmatter}

\section{Introduction}
In frame of General Relativity, black hole is a very attractive but
exotic object, generating extremely intense gravity to form an area
of ``trapped" surface \cite{Townsend}. The black hole attracts and
swallows all classical matter nearby, and it prevents them,
including light, from escaping. This implies black holes surrounded
by classical matter are hardly static, which is usually reported in
astrophysical observations. However, in a recent work \cite{bz},
Bronnikov and Zaslavskii found that some other kind of matter with
certain pressure and density may be in equilibrium with
4-dimensional black holes, which looks like the black holes have
surrounding ``hair".

It is an interesting work to study the black holes with surrounding
``hair'' not only in 4-dimensional spacetime, but also in higher
dimensions. As in string theory, higher-dimension is required to
avoid Weyl anomaly. For the bosonic string the critical dimension is
\(D_c=26\) and for superstring is \(D_c=10\) \cite{Polchinski}.
According to our analysis, the condition restricting the candidates
of the black hole ``hair''  such as cosmic string and quintessence,
also gives a constraint on the spacetime dimensions. The result
mainly based on General Relativity is consistent with
\cite{hellerman}, which comes from string theory. Furthermore, we
also find 4-dimensional spacetime is special.

Our letter will be arranged as follow, in the next Section we extend
the work of \cite{bz} to D-dimension, we obtain the same conclusion
as in 4-dimension. In Section 3, we consider concrete models as the
candidates of black hole hair: cosmic string, and two special
quintessence models. Analysis shows they are compatible with some
special choice of spacetime dimensions only. A general conclusion is
given in Section 4.

\section{Black hole hair in D-dimensional spacetime}
In the beginning, a general static, spherically symmetric metric in
D-dimensional spacetime writes
\begin{equation} \label{eq:1}
ds^2=A(u)dt^2-\frac{du^2}{A(u)}-r^2(u)d\Omega^2,
 \end{equation}
where \(\Omega\) is the area of an unit sphere in \(D-2\)
dimensions. The quasiglobal radial coordinate \(u\) is chosen to
ensure that it always takes a finite value \(u_h\) at a Killing
horizon where \(A(u)=0\) and to ensure analytic continuation of
the metric across the horizon \cite{bz}. Moreover, both \(A(u)\)
and \(r(u)\) should be analytic functions of \(u\) at \(u=u_h\). A
fine asymptotic behavior of \(A\) like \(A(u)\sim
(u-u_h)^n\)(\(n\in \mathbb{N}\)) assures a regular horizon. In the
case of a black hole, the solution of \(A(u)=0\) with maximum
\(u\) corresponds to the event horizon.

According to Eq. (1), we get the form of Ricci tensor
\begin{equation} \label{eq:2}
\begin{aligned}
& R^0_0=-\frac{A''}{2}-(D-2)\frac{A'r'}{2r},\\
& R^1_1=-\frac{A''}{2}-(D-2)(\frac{A'r'}{2r}+\frac{Ar''}{r}),\\
& R^i_i=-\frac{A'r'}{r}-\frac{Ar''}{r}-(D-3)\frac{1-Ar'^2}{r^2},
\end{aligned}
 \end{equation}
where \(i=3,4\dots D-1\) and the prime is a derivative with respect
to \(u\). Let us consider an ideal fluid as the source of
energy-momentum tensor \(T^\mu_\nu\) in Einstein equations
\(G^\nu_\mu\equiv R^\nu_\mu-\frac{1}{2}\delta^\nu_\mu R=-8\pi
T^\nu_\mu\),
\begin{equation} \label{eq:3}
T^\nu_\mu=Diag(\rho, -p_r, -p_\bot , -p_\bot \dots),
\end{equation}
where the density \(\rho\), the radial pressure \(p_r\) and the
transverse pressure \(p_\bot\) are functions of \(u\).

Firstly we obtain
\begin{equation} \label{eq:4}
G^0_0-G^1_1\equiv (D-2)A\frac{r''}{r}=-8\pi(\rho+p_r).
 \end{equation}
On one hand, if \(r''=0\) outside the event horizon, the only
possible solution is \(\rho=-p_r\). This solution can be interpreted
to cosmological constant or vacuum energy. This result is
semi-classical but natural. Although the energy of vacuum is
motivated in quantum theory and its asymptotic behavior near the
horizon is different with near infinity \cite{Townsend}, the black
hole is still static in frame of general relativity. An alternative
choice may be considered as a mixture of ordinary matter
(\(\rho+p_r>0\)) and phantom \((\rho+p_r<0)\). As we shall see in
the following, since phantom is excluded from black hole hair
candidate, it should not account for the mixture, either. On the
other hand, the \(r''\neq 0\) case leads more non-trivial solutions.
We just consider this case hereinafter.

The r.h.s of Eq.(4) vanishes near the horizon \(A(u_h)=0\), i.e.,
\begin{equation} \label{eq:5}
p_r(u_h)+\rho(u_h)=0.
 \end{equation}
We consider a kind of matter which has the linear equation of state
(EOS)
\begin{equation} \label{eq:6}
p_r=\omega\rho,\ \omega=const,
\end{equation}
both \(\rho\) and \(p_r\) should be smooth to 0 as \(u\rightarrow
u_h\). Therefore, we can expand \(A(u),\rho(u),r(u)\) with \(\Delta
u\equiv u-u_h\),
\begin{equation} \label{eq:7}
\begin{aligned}
& A(u)=A^n\Delta u^n[1+o(1)],\\
& \rho=\rho_0\Delta u^k[1+o(1)],\\
& r(u)=r_h+r'_h\Delta u+\frac{1}{2}r''_h\Delta u^2+o(\Delta u^2).
\end{aligned}
\end{equation}
where \(n\in \mathbb{N}\) is the order of the horizon, and \(k\in
\mathbb{N}\) is the number of the first non-vanishing term of the
Taylor series.

Matter near the horizon should at least satisfy the conservation law
\(\nabla _\nu T_\mu^\nu=0\). The only nontrivial term is
\begin{equation} \label{eq:8}
p'_r+(D-2)\frac{r'}{r}(p_r-p_\bot )+\frac{A'}{2A}(\rho+p_r)=0.
\end{equation}
As \(A\rightarrow 0\) near the horizon, with a weak assumption that
\(|p_\bot|/\rho<\infty\) , the term with \(r'\) in Eq. (8) is
neglected comparing with the third one \cite{bz} (moreover, in the
isotropic case, \(p_r=p_\bot \), this term will be vanishing), and
we obtain
\begin{equation} \label{eq:9}
\rho\sim A^{-(\omega+1)/(2\omega)},\ -1<\omega<0.
\end{equation}
which is the same result as in \cite{bz}, independent of D.
Combining this with equation (7), we get
\begin{equation} \label{eq:10}
\omega=-\frac{1}{1+2kn}.
\end{equation}
A discrete set of values belonging to (-1,0) of \(\omega\) is
allowed, while \(\omega\) of phantom does not sit in this interval.
Also, massive dust and radiations are excluded from the black hole
hair candidate immediately. It is consistent with our experience.

By using Eqs. (4) and (7) in the main approximation, it gives
\begin{equation} \label{eq:11}
(D-2)A_n\frac{r''_h}{r_h}\Delta u^n=-8\pi(\omega+1)\rho_0\Delta u^k,
\end{equation}
leading to the requirement \(k\ge n\), where \(k>n\) corresponds to
\(r''_h=0\). Taking the main approximation in \(G^1_1\),
\begin{equation} \label{eq:12}
G^1_1\equiv \frac{D-2}{2r^2}[A'r'r-(D-3)(1+Ar'^2)]=8\pi p_r,
\end{equation}
we have
\begin{equation} \label{eq:13}
-\frac{(D-2)(D-3)}{2}+\frac{D-2}{2}nA_nr_hr'_h\Delta u^{n-1}=8\pi
\omega \rho_0\Delta u^k.
\end{equation}
When \(u=u_h\), both sides of the equation should be 0. It requires
\(n=1\), and \(A_1r_hr'_h=D-3\). So we obtain \(\omega=-1/(1+2k)\),
 in which \(\omega=-1/3\) is the generic case with \(k=1\).
Moreover, if we consider a mixture of the candidate matter and a
vacuum fluid in the stress-energy tensor \cite{bz}
\begin{equation} \label{eq:14}
T^\nu_{\mu(vac)}=Diag(\rho_{(vac)}, -\rho_{(vac)}, -p_{\bot(vac)} ,
-p_{\bot(vac)} \dots).
\end{equation}
then \(n\ge 1\) is admissible.

\section{Candidates for black hole hair}

As we see in last section, the static and spherically symmetric
black hole in D-dimensional spacetime in Eq. (1) can be in
equilibrium with a non-interacting mixture of vacuum matter and
matter with \(\omega=-1/(1+2kn)\) near the horizon. It looks somehow
strange that \(\omega\) is independent of D. While in usual cases,
the EOS of matter always gets different while spacetime dimension
changes. Luckily, there exist two parameters for us to adjust
\(\omega\), so we might hope to get connections between k, n and D
in specific models\footnote{These models are usually required to be
isotropic. In fact, with a given vacuum fluid, it is not difficult
to find isotropic solutions of Eq. (1). One solution with
\(\rho_{vac}=const\) in 4-dimension is given in the example of
\cite{bz}}.

There are several methods to work out \(\omega\) of specific matter,
and here we choose a simple one. In the standard cosmological model,
the universe is usually described by the Robertson-Walker metric
\cite{Carroll}
\begin{equation} \label{eq:15}
ds^2=-dt^2+a^2(t)[\frac{dr^2}{1-Kr^2}+r^2d\Omega^2]
\end{equation}
where \(a(t)\) is scale factor with cosmic time t. The constant
\(K\) describes the geometry of the spatial section of spacetime,
with closed,flat and open universes corresponding to \(K=1,0,-1,\)
respectively. Consider an ideal perfect fluid as the source of the
energy momentum tensor \(T^\mu_\nu=Diag(\rho,p,p,p\dots)\). From
Einstein equations, we can get two independent equations
\begin{equation} \label{eq:16}
\begin{aligned}
&H^2\equiv \ (\frac{\dot{a}}{a})^2=\frac{16\pi
G\rho}{(D-1)(D-2)}-\frac{K}{a^2},\\
&\dot{H}=-\frac{8\pi G(\rho+p)}{D-2}+\frac{K}{a^2}.
\end{aligned}
\end{equation}
They lead to the continuity equation
\begin{equation}\label{eq:17}
\dot{\rho}+(D-1)H(\rho+p)=\dot{\rho}+(D-1)(1+\omega)(\frac{\dot{a}}{a})\rho=0.
\end{equation}
After a simple calculation, we obtain
\begin{equation}\label{eq:18}
\rho\sim a^{-(D-1)(1+\omega)}.
\end{equation}
So we just need to know the power relation between the energy
density \(\rho\) and the scale factor \(a\), then the equation of
state \(\omega\) can be easily derived. For example, the baryonic
matter has \(\rho\sim a^{-(D-1)}\), corresponding to \(\omega=0\).

Among all the discrete values of Eq. (10), \(-1/3\) is quite special
in 4-dimensional spacetime, which can be seen in many cosmological
literatures. It's a critical number in 4-dimensional FRW spacetime.
We get an accelerated expanding universe when \(\omega<-1/3\) and a
decelerating universe when \(\omega>-1/3\). It also corresponds to
the EOS of ``curvature" in FRW universe while \(K\neq 0\)(Usually it
is useful to pretend the energy density of spacial curvature to be
\(\rho_k \sim k/a^2\) and \(\omega_k=-1/3\) in 4-dimension). So
hereafter we require that the black hole hair takes \(\omega=-1/3\)
when \(D=4\), and allows other values as D changes.

\subsection{Cosmic strings}

Cosmic strings first emerged from quantum field theory in early
1980s \cite{Vilenkin}. A cosmic string is a hypothetical
1-dimensional macroscopic topological defect in spacetime. These
strings might produce disturbances that led to the structure in our
universe. More interestingly, in recent work, Polchinski showed that
many of the properties of cosmic strings can be used to investigate
superstring theory, especially superstrings of cosmic length
\cite{Polchinski2}.

Since cosmic string is a 1-dimensional object, the total energy
grows as its length \(a\). So the density scales as \cite{Shellard}
\begin{equation}\label{eq:19}
\rho_s\sim\frac{a}{a^{D-1}}\sim a^{2-D}.
\end{equation}
Comparing (19) with (18), we can get
\begin{equation}\label{eq:20}
\omega_s=-\frac{1}{D-1},
\end{equation}
which is -1/3 in 4-dimension spacetime.

However, when \(D\neq 4\), it is obviously that \(\omega_s\neq
-\frac{1}{3}\). In order to make cosmic strings the black hole hair,
we have \(-\frac{1}{D-1}=-\frac{1}{1+2kn}\), and get
\begin{equation}\label{eq:21}
D=2+2kn.
\end{equation}
We analyze this result as follows.

On one hand, \(D\) must be even. In fact, a similar result has been
found in \cite{hellerman}. In that work, Hellerman and Swanson
analyzed the dimension changing in string theory through closed
string tachyon condensation. They found that in the superstring
case, the number of spacetime dimensions is reduced by \(\Delta
D=2\), in order to preserve the chiral R-parity. It implies that
\(D\) should keep its parity while dimension changes. Since we live
in 4-dimensional universe, the nature number of spacetime dimensions
should be even, as we assert here. However, our solution is mainly
based on General Relativity.

On the other hand, as \(k,n>0\), according to Eq. (21) we can find
that \(D\) should not be less than \( 4\). Then the 3-dimensional
spacetime is rejected\footnote{With a simple calculation, one can
find that in 2-dimension \(G^\mu _\nu\equiv 0\). So considering
black hole hair in 2-dimensional spacetime makes no sense.}. There
are many ways to lower the number of spacetime dimensions down, such
as by using compactification and tachyon condensation, however, few
of these methods implicate the lower bound four.

\subsection{Quintessence}

Today, a wide variety of scalar field models have been proposed to
describe the dark energy. Quintessence model is a popular one of
them \cite{Copeland}. Quintessence, as an alternative to a positive
cosmological constant, can realize a fluid with a equation of state
\(\omega>-1\). A general action for quintessence in D-dimensional
spacetime is given by
\begin{equation}\label{eq:22}
S=\int\d^Dx\sqrt{-g} [-\frac{1}{2}(\nabla \phi)^2-V(\phi)].
\end{equation}
where \(V(\phi)\) is the potential of the field. Here we use
\(V(\phi)=c\exp(\gamma\phi)\), with \(c,\gamma>0\). A calculation in
\cite{hellerman2} shows in this case\footnote{Here we just consider
a flat universe(\(K=0\)) for simple. In fact, observations have
shown that the current universe in very close to a spatially flat
geometry}, \(\omega\) takes
\begin{equation}\label{eq:23}
\omega=-1+\frac{(D-2)\gamma^2}{2(D-1)}.
\end{equation}

If \(\gamma\) is a constant, in order to get \(\omega=-1/3\) when
\(D=4\), we shall have \(\gamma=\sqrt2\). Substituting this to the
D-dimensional case, we get
\begin{displaymath}
\omega=-\frac{1}{D-1}.
\end{displaymath}
which is the same as we get in Eq. (20). However, this quintessence
model is quite different with cosmic string.

Furthermore, linear dilaton model in string theory can be also
connected to quintessence \cite{hellerman2}. The effective spacetime
action of a linear dilaton CFT is
\begin{equation}\label{eq:24}
S=\frac{1}{2\kappa^2}\int\d^DX \sqrt{-G}
[-\frac{2(D-D_c)}{3\alpha'}e^{2\phi/\sqrt{D-2}}-(\nabla\phi)^2].
\end{equation}
where \(\phi\) is the dilaton field. Comparing Eq. (24) with (22),
we can find \(\gamma=2/(D-2)\) and then
\begin{equation}\label{eq:25}
\omega=-\frac{D-3}{D-1}.
\end{equation}
which is \(-1/3\) when \(D=4\). However, this solution can not be
equal to \(-1/(1+2kn)\) when \(D\) changes. So here 4-dimensional
spacetime become special again. Only in this case the linear dilaton
string model can be a candidate for black hole hair.

\section{Conclusions and Outlook}

In conclusion, we have investigated the black hole hair in
D-dimension spacetime, and found out its EOS satisfies
\(\omega=-1/1+2kn\), irrelative with \(D\). Analyzing cosmic string
as a candidate of black hole hair implicated that, (i)only even
number of spacetime dimensions is allowed. This result is also
consistent with \cite{hellerman}, it might reveal the relation of
cosmic string and superstring in an other aspect. (ii)4-dimension is
the lowest case of spacetime dimension. We also analyzed two special
models of quintessence. One got the same result as cosmic string.
The other showed the static black hole could be surrounded by linear
dilation field(in string theory) only in 4-dimension.

In context, we have assumed \(\omega\) of the black hole hair
candidate to be \(-1/3\) in 4-dimension. However, one can give out
other models of the candidate in 4-dimension, e.g.
\(\omega=-1/1+2kn\) with \(kn \ne 1\). It may be interesting to
reveal the physical sense of these models and see whether they can
give a limit on the spacetime dimensions. What's more, in cosmology,
it is useful to pretend the spacial curvature in FRW metric to be a
kind of matter, with \(\rho_k \sim k/a^2\) and
\(\omega_k=-(D-3)/(D-1)\) \cite{Carroll}. However, there is no
conservation equation for spacial curvature, we can not work on it
directly. A strict method is required to investigate on the relation
between curvature and black hole hair. Finally, it is also
interesting to generalize the analysis to non-spherical and rotating
distributions of matter.
\section*{Acknowledgements} We would like to thank Y. Xiao and L.Chen for useful discussions. This work is supported in part by the NNSF of China
Grant No. 90503009, No. 10775116, and 973 Program Grant No.
2005CB724508.

\end{document}